# Data-driven exploration of the strength-ductility synergy in refractory high-entropy alloys

Feng Zhao[a,b], Jingya Zhang[a], Yin Zhang[a,b,*], Huiling Duan[a,b,*]

[a] *State Key Laboratory for Turbulence and Complex Systems, Department of Mechanics, School of Mechanics and Engineering Science, Peking University, Beijing 100871, China*

[b] *HEDPS, CAPT and IFSA, Collaborative Innovation Center of MoE, Peking University, Beijing 100871, China*

## Abstract

Refractory high-entropy alloys (RHEAs) have attracted increasing attention for their exceptional ability to retain strength at high temperatures. However, current research mainly concentrates on compositions close to equimolar alloys, leaving a large part of the composition space unexplored. Here, we develop a multiscale framework based on a data-driven "dislocation genome" method that constructs energy landscapes for edge dislocation motion across various compositions. The local configurational force derived from the energy landscape is incorporated into dislocation dynamics simulations, revealing a scaling law connecting macroscale strength and atomic-scale local resistance governed by fluctuations in the local configurational force. Based on this scaling law, a new solid-solution strengthening model for edge dislocation in BCC alloys is developed, followed by a data-driven exploration of refractory alloys considering both strength and ductility. The developed method and theory are adaptable, allowing for extension to other BCC alloy systems and guiding high-throughput alloy design.

*Corresponding authors.

E-mail addresses: yinzhang@pku.edu.cn (Y. Zhang), hlduan@pku.edu.cn (H. Duan)

## Introduction

Refractory high-entropy alloys (RHEAs) have emerged as promising candidates for high-temperature applications such as turbine blades(*1*) and first wall materials in fusion reactors(*2*), owing to their exceptional mechanical performance under extreme conditions(*3-6*). The major challenge in the design of RHEAs currently lies in balancing high-temperature strength with room-temperature ductility: compositions with superior high-temperature strength exhibit poor ductility at room temperature, while improvements in ductility through compositional adjustments often compromise high-temperature strength(*7-9*). Understanding the unique dislocation behavior in RHEAs is crucial for alloy composition optimization(*10*). Previous studies have demonstrated that the localized pinning and depinning behavior of edge dislocations is critical to the synergy between strength and ductility in high-entropy alloys(*11, 12*). However, a systematic understanding of the cause of pinning, the morphological evolution of dislocations induced by pinning, and the associated dislocation dynamics remains lacking.

Compared to BCC systems, dislocation pinning in face-centered cubic (FCC) high-entropy alloys has been explored more extensively. Prior works developed a pinning criterion based on localized generalized stacking fault energy (GSFE), which successfully predicted pinning sites consistent with molecular dynamics simulations(*12-15*). A further study incorporates the effects of short-range order (SRO) on dislocation pinning(*16-18*). The validity of these criteria stems from the intrinsic dislocation core structure in the FCC lattice, i.e., dislocations split into widely spaced partial dislocations separated by stacking fault ribbons. Therefore, pinning strength depends not only on dislocation-solute interactions but also on the energy barriers for stacking fault ribbon motion, which are directly reflected by local GSFE variations. In contrast, dislocations in the BCC lattice have a compact core structure without dissociation into partials or forming stacking fault ribbons, rendering GSFE-based criteria inapplicable(*19*). Given the importance of understanding solid-solution strengthening mechanisms, the nano-pinning mechanisms for BCC edge dislocations should be further explored.

Recently, data-driven approaches have gained increasing attention to explore the compositional space of BCC alloys. Several studies directly utilize experimental data and machine learning models to design BCC alloys with target properties(*20-22*). However, these approaches are often limited by uncalibrated experimental bias and insufficient physical interpretability. Alternatively, other works attempt to build material property databases by employing methods such as *ab initio* simulations (*23, 24*). While these methods incorporate physical mechanisms to some extent, plastic deformation inherently involves multiscale correlations. Solely relying on microscopic properties makes it difficult to achieve reliable predictions of macroscopic performance. Furthermore, current research has rarely explored multi-objective optimization problems. Addressing trade-offs, such as the strength-ductility trade-off, still largely relies on empirical adjustments.

Here, we introduce a data-driven framework for exploring the compositional space of BCC alloys through an integrated multiscale modeling approach. We construct the realistic energy landscape of edge dislocation glide by combining large-scale atomistic simulations with neural networks (NN), and systematically analyze the nano-pinning of edge dislocations. We further incorporate this solute-induced energy landscape into a stochastic discrete dislocation dynamics (DDD) model, enabling the mesoscale simulation of the edge dislocation up to microseconds. From the DDD simulations, we reveal a scaling law connecting the solid-solution strengthening and microscale property fluctuation, which in turn informs a composition- and temperature-dependent solid-solution strengthening model of edge dislocations. This model allows for the prediction of high-temperature strength across the NbMoTaW compositional space. Coupled with a ductility model based on relative mobility between screw and edge dislocations, we conduct multi-objective optimization to balance high-temperature strength and room-temperature ductility, thereby facilitating the synergistic design of RHEAs.

## Results

### Data-driven construction of energy landscape

According to the dislocation mechanics, the total energy of a dislocation line is composed of three parts: the potential energy, the line tension energy, and the work done by external loading. For edge dislocations in BCC alloys, the potential energy is mainly contributed from the solute-dislocation interaction, because the intrinsic lattice resistance of edge dislocation motion in a perfect BCC lattice is very low. Therefore, the resistance of edge dislocation motion in BCC alloys is determined by solid-solution strengthening, which is characterized by the solute-dislocation interaction energy.

As demonstrated in our previous work(*25*), the solid-solution strengthening in the NbMoTaW alloy is dominated by the dislocation core region, which can give a good approximation of the interaction energy $E_{\mathrm{int}}$. Based on this physical picture, the solid-solution strengthening can be attributed primarily to atomic configurations near the slip plane, rather than to the countless atoms throughout the crystal. This locality significantly simplifies the problem, enabling quantitative characterization of the energy landscape of dislocation glide. Drawing an analogy to the concept of genes in biology, we regard the atomic arrangement in HEA as a form of "DNA sequencing" that encodes the energy landscape, and the local atomic arrangement near the slip plane as the "genetic determinant", which governs dislocation slip behavior. Based on this concept, we develop a data-driven model to construct the energy landscape from actual atomic configurations, termed the "dislocation genome" method, which is illustrated in Fig. 1(A). The dimensions of local atomic configurations considered and the method for calculating the interaction energies are elaborated as follows. Firstly, the dislocation slip plane is divided into minimal constituent units according to the periodicity of the BCC crystal. Along the slip plane's normal direction, six atomic layers of (110) planes adjacent to the slip plane are adopted. The minimal periodic unit along dislocation length corresponds to one [112] vector dimension, while the localized region along slip direction should encompass the dislocation core width, approximately $5b$. Thus, the atomic layers near the slip plane in Fig. 1(A) can be discretized into voxels of the dimensions above. Following this discretization, a neural network is constructed to identify atomic configurations within these voxels and output interaction energies across different positions, thereby establishing an energy landscape as illustrated in Fig. 1(A). Training data for neural networks is generated through an embedding model: local atomic configurations are extracted from NbMoTaW alloy and embedded into a pure metal matrix. (Details of the embedding procedure and the simulation configuration are shown in Fig. S1.) The interaction energy is calculated as the core energy difference between edge dislocation segments in alloy atomic configurations $E_{\text{core}}^{\text{alloy}}$ and those in pure metal $E_{\text{core}}^{\text{pure}}$, i.e. $E_{\text{int}} = E_{\text{core}}^{\text{alloy}} - E_{\text{core}}^{\text{pure}}$. Notably, although dislocation energies vary among pure BCC metals, their intrinsic Peierls barriers are negligible compared to the interaction energy variation arising from solid-solution strengthening. Hence, the choice of matrix element has little effect on the overall shape of the energy landscape. Here, we choose Nb as the reference matrix in calculations. Furthermore, matrix atoms distant from dislocation cores induce negligible solid-solution strengthening, so the effect of these local configurations can be evaluated independently. The corresponding interaction energy as a function of the edge dislocation's position is then computed via the Nudged Elastic Band (NEB) method(*26*).

The computed interaction energy $E_{int}$ in each voxel is aligned to construct the entire energy landscape on the slip plane. This alignment along the dislocation slip direction is verified in Fig. 1b, taking the equimolar NbMoTaW alloy as an example. The energy values assembled from the minimum energy path (MEP) of the embedding model coincide with those from full NEB calculations, confirming that the interaction energy can be superimposed along the slip direction. Analogously, the validation of energy additivity along the dislocation line direction is shown in Fig. S2. Figure 1(C) displays the energy prediction of NN on training, test, and validation datasets. Notably, despite being primarily trained on equimolar compositions, the NN achieves accurate predictions for non-equimolar alloys in the validation set, demonstrating the model's scalability and capability to explore the entire compositional space of NbMoTaW alloys.

The "dislocation genome" method derives the energy landscape governing dislocation motion based on atomic configurations. While primarily targeting edge dislocations in BCC alloys, where their energy landscapes are far more complex than screw dislocations and dominate high-temperature strength, this method is generalizable. In essence, the method's central idea lies in decomposing crystalline defects into fundamental units, analyzing their interactions with alloying atoms, and constructing energy landscapes to predict the evolution of defect systems. This framework can be extended to screw dislocations in BCC alloys, dislocations in other crystal structures (e.g., FCC or HCP), or even other defect types, such as grain boundaries, twin boundaries, and stacking faults.

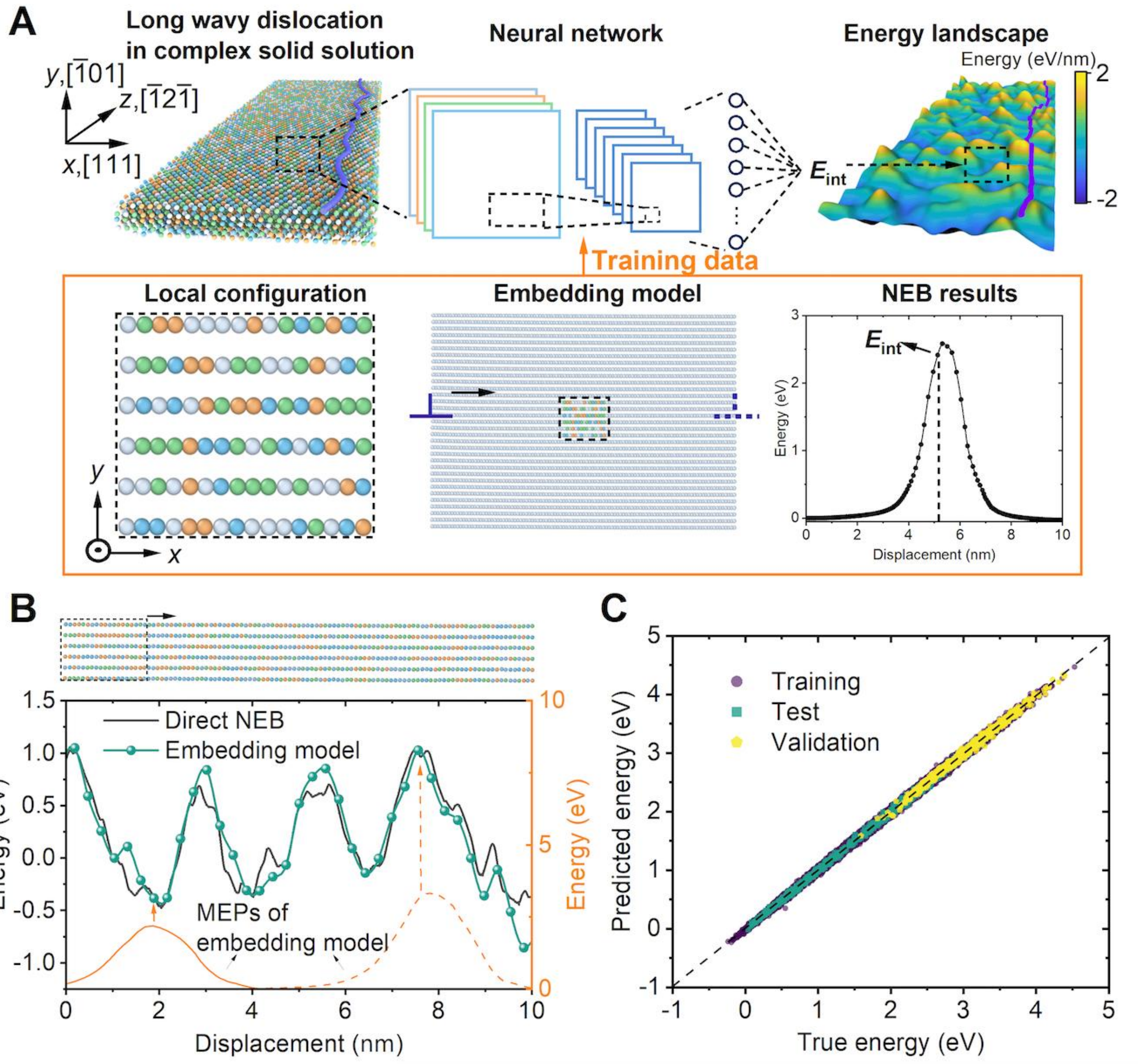

**Fig. 1 Data-driven "dislocation genome" model development and evaluation. (A)** Flow chart of generating the energy landscape from the atomic configuration. **(B)** Superposition of embedding model results and comparison with direct NEB simulations. **(C)** Predicted local energy for training, test, and validation set.

## Composition effect on the energy and force landscapes

Building on the established "dislocation genome" method, we investigate the characteristics of edge dislocation energy landscapes and the corresponding configurational forces for various compositions within the NbMoTaW system. The concept of configurational force, initially proposed by Eshelby, refers to a fictitious force acting on defects, obtained by differentiating the total energy with respect to defect position. The local configurational force imposed on a dislocation can be derived from the dislocation-solute interaction energy, i.e., $f_{\mathrm{int}}=-\partial E_{\mathrm{int}}/\partial x$. Drawing parallels with the Peach-Koehler force expression for dislocations, $f_{\mathrm{int}}$ here can also be expressed as the product of a stress-dimensioned quantity and the Burgers vector, i.e., $f_{\mathrm{int}} =\tau_{\mathrm{int}}b$, where $\tau_{\mathrm{int}}$ is the configurational stress. Since the Burgers vector length $b$ in the studied alloy system varies minimally, which ranges between 0.27 to 0.28 nm, $\tau_{\mathrm{int}}$ can be used to represent the magnitude of configurational force $f_{\mathrm{int}}$. Figure 2 illustrates the effects of different quaternary alloy compositions on the energy landscape and configurational force for a 25 nm × 25 nm slip plane. As shown in Fig. 2(A), the distribution of local interaction energy follows a Gaussian profile. The $Nb_{10}Mo_{10}Mo_{68}W_{12}$ alloy exhibits the smallest variance, while $Nb_{45}Mo_{5}Mo_{5}W_{45}$ shows the largest variance, with the equimolar alloy falling in between. This result suggests that while higher configurational entropy generally implies greater local property fluctuations, the equimolar alloy with the maximal configurational entropy does not necessarily exhibit the largest fluctuations of dislocation resistance. The variance manifests visually as the undulation intensity of the energy landscape, with contour maps shown in Fig. 2(B)-(C). Distributions of the configurational force derived from finite differences of the energy landscape, and the contour maps are presented in Fig. 2(D)-(F).

Qualitatively, the energy landscape shows minimal significant in-plane anisotropy, whereas configurational forces display clear directional dependence, with strong short-range correlations observed along the *x*-direction. To quantify these spatial correlations, we perform an autocorrelation analysis. The autocorrelation coefficient, a dimensionless quantity ranging from -1 to 1, measures the extent of spatial dependency: values near 1 indicate strong positive correlation, values near -1 correspond to strong negative correlation, and values around 0 suggest no spatial correlation. Fig. 2(G)-(H) reveals that atomic configurations within 1.5 nm exhibit correlated energy and force values, but lose correlation beyond this distance, consistent with a random solid solution lacking SRO. Notably, energy correlations decay monotonically to zero, while force correlations first drop to -0.4, rebound to zero, and then fluctuate weakly, corresponding to a random force disorder correlator(*27*). This correlation explains the discrepancy between energy and force contour maps in Fig. 2. It should be noted that this study focuses on random solid solutions. Given the widespread interest in SRO within high-entropy alloys, spatial correlations that evolve in systems with varying degrees of SRO should be addressed in future work.

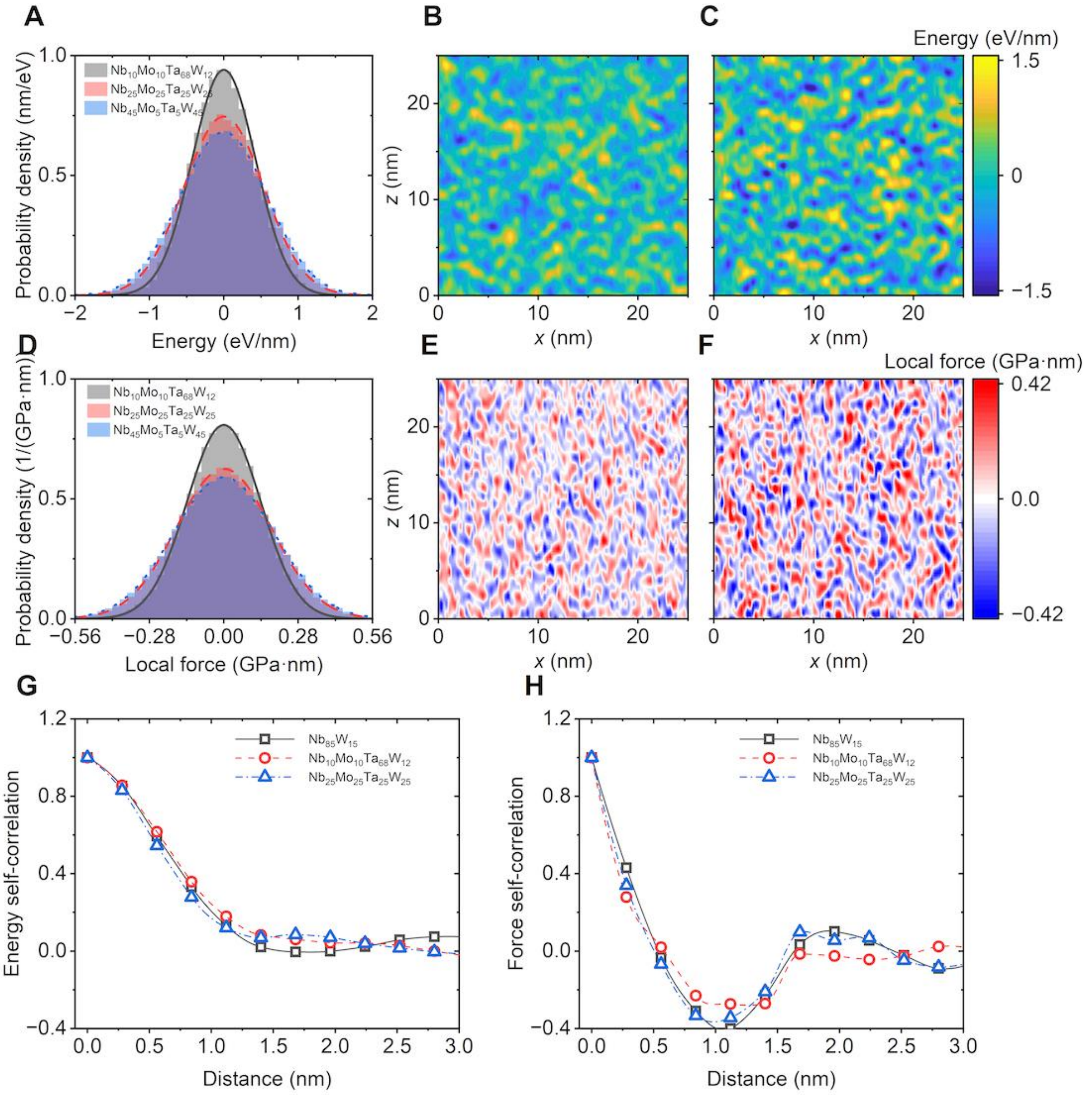

**Fig. 2 Effect of alloy compositions on the energy landscape and force landscape. (A)** Probability distribution of local interaction energy, contour plot of two compositions in **(B)** and **(C)**. **(D)** Probability distribution of configurational force, contour plot of two compositions in **(E)** and **(F)**. Normalized auto-correlation of energy **(H)** and force **(I)**.

**Validation of the neural network-predicted energy landscape**

Direct NEB simulations of long dislocation segment glide yield the MEPs and reveal nanoscale pinning and depinning processes. By comparing the resulting dislocation morphologies and pinning sites with the energy landscape predicted by the “dislocation genome” method, deeper insights into nanoscale pinning mechanisms can be achieved. Results are now presented for both binary NbW alloys and quaternary NbMoTaW alloys. Fig. 3(A)-(B) illustrates the correspondence between the MEP of long-distance dislocation glide in NbMoTaW, dislocation morphologies, and the energy landscapes predicted. The case of NbW alloy is shown in Fig. 3(C)-(D). The MEP and dislocation configurations were obtained through large-scale NEB calculations, with a dislocation segment length of 30 nm. The simulation employed 2,500 replicas to discretize the 40 nm glide distance, with each replica containing approximately 1.2 million atoms, collectively representing a simulation involving 0.3 billion atoms. Local minima along the MEP, corresponding to dislocation pinning sites, are marked by black dots in Fig. 3(A). The dislocation line shapes at these positions are sequentially overlaid with the neural network-predicted energy landscape in Fig. 3(B). Yellow regions denote high-energy “peaks” that hinder dislocation motion, while dark blue regions represent “valleys” where pinning is favored. Pinning sites are highlighted with white circles. The dislocation is observed to pin in deep valleys, which is illustrated in the enlarged three-dimensional plot in Fig. 3(E), corresponding to the region marked by dashed rectangle in Fig. 3b. The dislocation can also be caught between two peaks, while its motion involves locally overcoming one or multiple peaks, as shown in Fig. 3(F)-(K), a magnified view of Fig. 3(D). In Fig. 3(F)-(K), dashed lines indicate prior configurations to emphasize positional evolution. Starting from the initial state (Fig. 3(F)), subsequent panels illustrate five incremental motion steps, each involving the localized overcoming of one or multiple energy peaks. The above results confirm that the predicted energy landscapes effectively explain the nano-pinning mechanism of BCC edge dislocation.

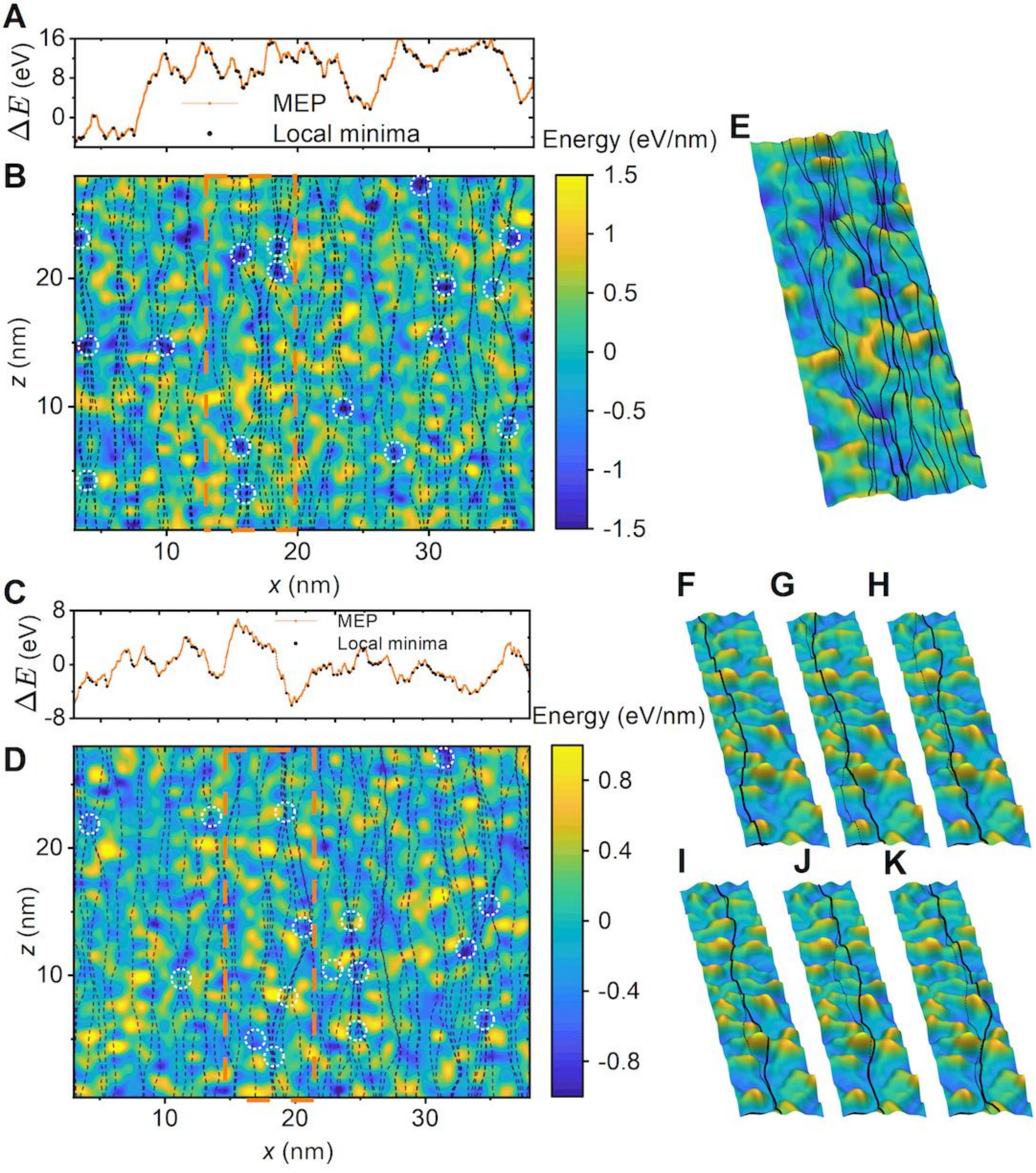

**Fig. 3 Comparison of energy landscape predicted by neural networks and direct NEB simulations of long dislocation segments. (A)** MEP of NEB simulations with a dislocation length

of about 30 nm, in which the local minima are marked. **(B)** Neural network predicted energy landscape superposed with pinned dislocation configurations corresponding to the local minima in **(A)**. **(C)** MEP of NEB simulations, in which the local minima are marked. **(D)** Neural network predicted energy landscape superposed with pinned dislocation configurations corresponding to the local minima in **(C)**. **(E)** Enlarged three-dimensional plot of the framed part in b, showing the pinning of dislocations in deep local minima. **(F)-(K)** Three-dimensional plot of the framed part in **d**, showing the successive overcoming of local barriers.

**Temperature-dependent critical shear stress obtained by stochastic DDD simulations**

While solid-solution strengthening gives rise to configurational forces, these forces do not necessarily manifest as local resistance. As shown in Fig. 2, these forces exhibit a zero-mean Gaussian distribution, implying equal probabilities of locally resistive or propulsive effects on dislocation segments. However, such a distribution does not cancel out macroscopically; instead, it collectively acts as resistance, highlighting the importance of studying its mesoscale impact. Simulating the motion of a long dislocation line using the NEB method requires substantial computational resources, highlighting the need for a more efficient approach.

To overcome these limitations, we develop a discrete dislocation dynamics (DDD) framework capable of modeling mesoscale dislocation motion incorporating both thermal fluctuation and local configurational forces. In this framework, random forces derived from the thermal fluctuations are represented using a Langevin-based model, extending the classical Langevin equation from Brownian motion to model crystalline defects, enabling parameter-free prediction of the transition from the thermo-activated behavior to phonon drag behavior. While Langevin-based approaches have been applied in previous studies, we integrate the local configurational force into the framework to simulate the motion of dislocations across the complex energy landscapes of HEAs. (see Figs. S3 and 5 for the simulation flowchart and the DDD simulation geometry).

To compare with experimental data(28-30), in Fig. 4(A)-(H), the critical stress for edge dislocation motion is extracted from stress-strain rate curves. Theoretically, the critical resolved shear stress corresponds to quasi-static loading at experimental strain rates (~$10^{-3}$ $s^{-1}$). However, due to simulation time limitations, here we adopt a threshold strain rate of 1 $s^{-1}$ as an approximation (see Fig. S9). The temperature-dependent critical stress is fitted with a power-law formula,

$$\tau_c(T) = \tau_0 \left( 1 - \left( \frac{T}{T_0} \right)^p \right), \quad (1)$$

where $\tau_0$ denotes the critical shear stress at 0 K, $T_0$ denotes the temperature at which the critical shear stress for edge dislocation motion is zero. The two parameters $\tau_0$ and $T_0$ are related to alloy composition. Considering the fitting performance, the optimal exponent $p$ is determined as 0.4. Using this $p$ value, other parameters are derived. Notably, the fitted $T_0$ exceeds alloy melting points (e.g., $T_0$ = 2500 K for $NbW_{0.01}$, slightly above Nb's melting point). Thus, $T_0$ holds theoretical significance only. Near melting temperatures, mechanisms like grain boundary sliding or dynamic recrystallization dominate, rendering dislocation-slip theories inapplicable. Figure 4(I) compares simulated and experimental critical shear stresses across various BCC alloys. The simulation results match well with experimental data, validating the constructed energy landscape and the new DDD framework.

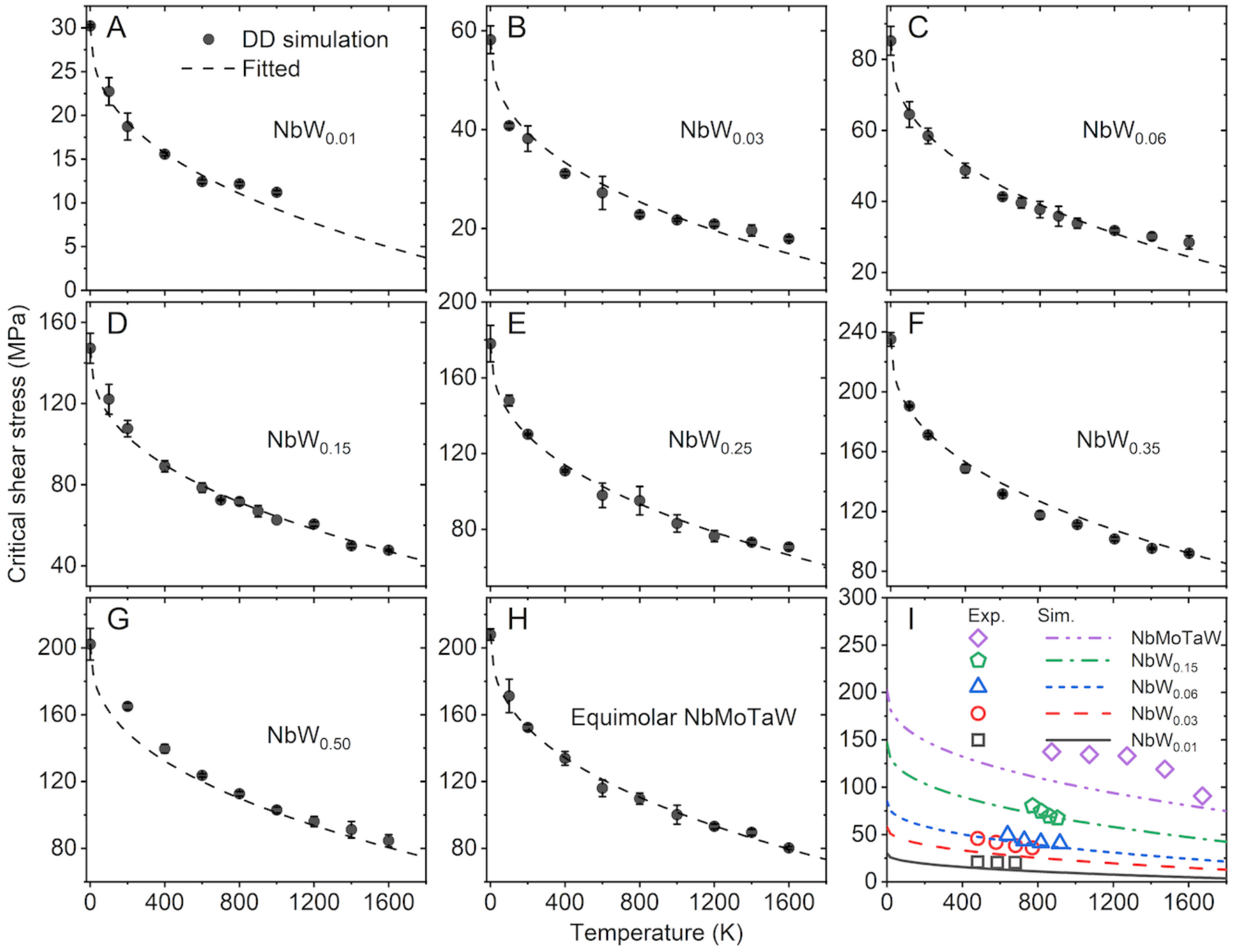

**Fig. 4 Critical shear stress at different temperatures obtained by DDD and comparison with experimental data. (A)-(G)** Critical shear stress at 0 – 1600 K of NbW refractory alloys with varying W concentration. **(H)** Simulation results of equimolar NbMoTaW alloy. **(I)** comparisons with experimental data.

## Scaling law between critical stress and local force fluctuation

To further analyze the DDD results, we define the fluctuation of local configurational force as

$$\tau_{\mathrm{std}} = \frac{1}{b}\sqrt{\left\langle f_{\mathrm{int}}^{2} \right\rangle}\,, \tag{2}$$

considering the mean of local configurational force is 0. $\tau_{\mathrm{std}}$ is affected by the alloy composition. Figure 5(A) shows the relationship between the critical shear stress at 0 K $\tau_0$ and the fluctuation of local configurational force $\tau_{\mathrm{std}}$, which corresponds respectively to the macroscopic effect of solid-solution strengthening and the microscopic fluctuation. A linear correlation is found between $\tau_0$ and $\tau_{std}$, i.e.,

$$\tau_0 = \xi \tau_{std}, \tag{3}$$

where $\xi$ is the slope. This scaling law reveals the cross-scale correlation between the local composition fluctuation in BCC alloys and the solid-solution strengthening. Considering the extreme case of pure BCC metals, the corresponding $\tau_{std}$ is zero, and thus $\tau_0$ is zero as predicted by this scaling law. This conforms to the near-zero resistance of edge dislocation in pure BCC metals. The scaling law can be rationalized by the Π theorem. For a dislocation segment with unit length, the equation of motion is $m\dot{v}=-Bv+\tau_{int}b\varsigma+\Gamma d^2x/dz^2+\tau_{ext}b$, where $\tau_{ext}$ is the external loading. As mentioned before, the local stress $\tau_{int}$ conforms to a normal distribution with a mean value of zero; it can be expressed as the multiplication of the standard deviation $\tau_{std}$ and a dimensionless random number $\varsigma \sim N(0,1)$. Therefore, the equation of motion can be rewritten as $m\dot{v}=-Bv+\tau_{std}b\varsigma+\Gamma d^2x/dz^2+\tau_{std}b$. According to the Π theorem, the critical external loading $\tau_c$ satisfies the following relationship with other physical quantities: $\tau_c/\tau_{std} = \Psi(\Gamma/\tau_{std}b))$, where Ψ is an undetermined function. If Ψ is a constant function, then there is a linear relationship between $\tau_c$ and $\tau_{std}$.

The temperature parameter $T_0$ also has a linear relationship with $\tau_{std}$ as shown in Fig. 5(B), which can be expressed as $T_0(\tau_{std})= T_b+\zeta\tau_{std}$. Fig. 5(C) illustrates the relationship between the critical shear stress at 0 K (see Fig. S10) and W concentration in binary NbW alloys. A power-law fit to the simulation results reveals an approximate square-root dependence of the critical shear stress on W concentration, consistent with the classical strong-pinning model, the Friedel model. Furthermore, the DDD results align with atomic-scale simulations. The above results lead to a

composition- and temperature-dependent model for the solid-solution strengthening of BCC edge dislocations, i.e.,

$$\tau_c(T) = \tau_0(\tau_{std})\left(1-\left(\frac{T}{T_0(\tau_{std})}\right)^p\right), \tag{4}$$

where the composition dependence is reflected by $\tau_{std}$. Specifically, the "dislocation genome" method can be applied to analyze slip planes in target compositions and $\tau_{std}$. By integrating this theoretical framework with the "dislocation genome" method, high-temperature performance predictions for BCC alloys become feasible. As shown in Fig. 5(D), the high-temperature yield strength of alloys within the NbMoTaW system is compared with model prediction, including $TaW_{0.10}$(*31*), $NbMo_{0.05}$(*32*), $NbW_{0.056}$(*28*), $NbTa_{0.28}W_{0.10}$(*33*), and $TaW_{0.04}$(*34*), all of which fall within the 95% prediction interval. It should be noted that the current model (Eq. (4)) considers only the resistance for edge motion in single-phase BCC random solid solutions. In real materials, various strengthening mechanisms such as SRO, secondary phases, and grain-boundary effects come into play. Since the proposed random solid-solution model operates at a finer scale, it can serve both as a foundation for analyzing larger-scale mechanisms and as a modular component for crystal plasticity simulations. For example, the critical resolved shear stress derived here can be incorporated into crystal plasticity frameworks superposed with other strengthening contributions.

Several parameters reflecting microscale property fluctuations have been proposed in existing studies on HEAs, such as variations in generalized stacking fault energy, local atomic stress fluctuation, lattice distortion parameter, and size mismatch parameter. Fig. 5(E) illustrates the relationship between $\tau_{std}$ and the atomic size mismatch parameter $\delta = \sqrt{\sum_i c_i\left(1-\frac{r_i}{\bar{r}}\right)^2}$ with $\bar{r} = \sum_i c_i r_i$ , while Fig. 5(F) compares the local stress fluctuation calculated from the CNN-predicted energy landscape with the atomic mean square shear stress obtained from Rodney's microelasticity model(*35*) (see Supplementary Information for details). In the composition range approaching binary "solute-solvent" alloys, both quantities exhibit an approximately linear relationship. However, in the near-equiatomic region, the two measures deviate significantly, which indicates that the conventional misfit parameters cannot fully capture the mixing effect on dislocations in HEAs. Moreover, these parameters (e.g., size misfit or atomic mean square shear stress) characterize defect-free alloy systems where atoms reside in equilibrium states. In contrast, local configurational force fluctuations $\tau_{std}$ proposed in this work lie in the explicit consideration of dislocations within alloy lattices, effectively capturing the non-equilibrium nature that dislocations induce within their core regions.

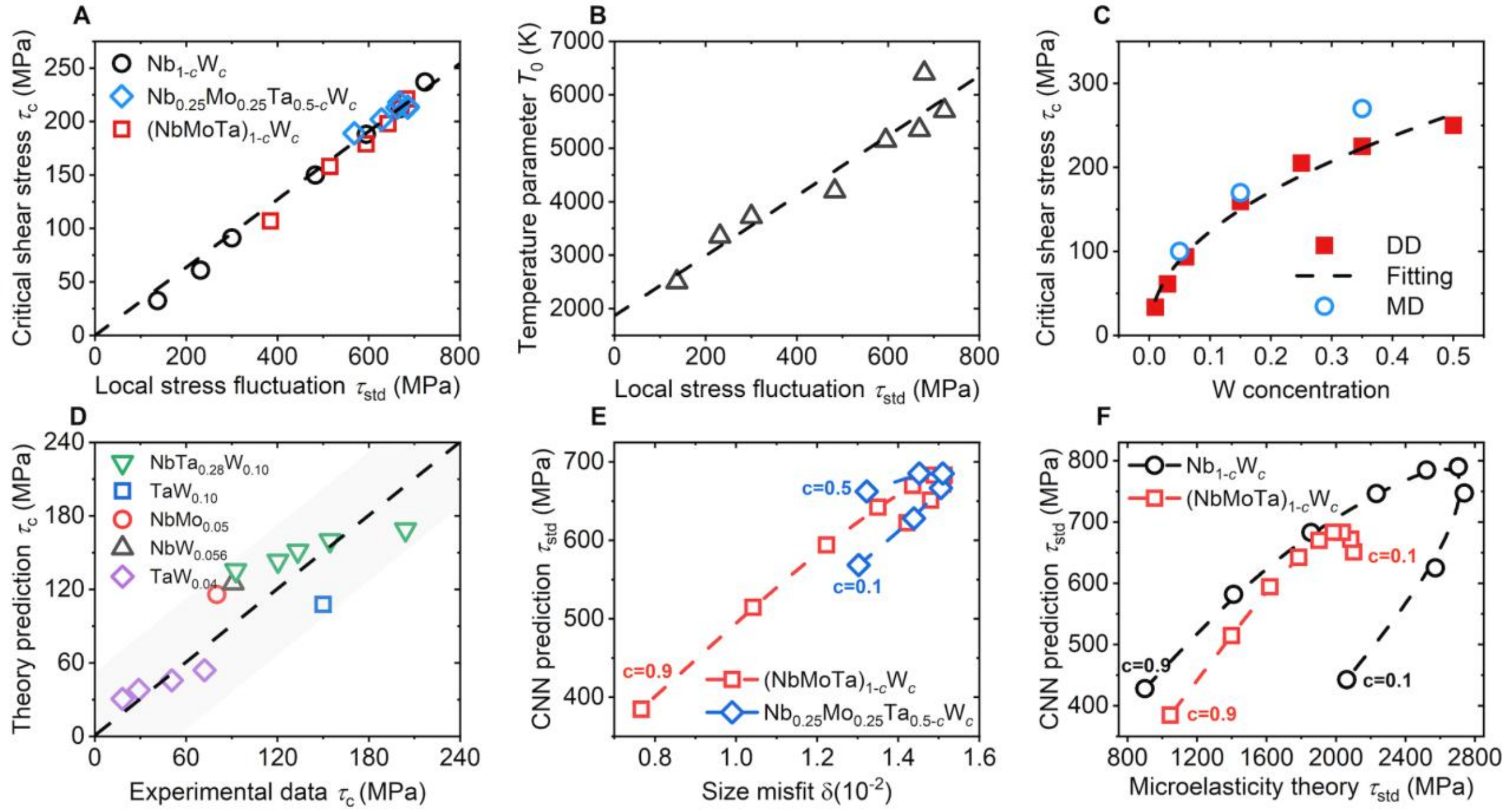

**Fig. 5 Scaling law between critical shear stress and the fluctuation of local configurational force, and validations. (A)** Linear relationship between athermal strength and the fluctuation of configurational force $\tau_{std}$. **(B)** Linear relationship between the parameter $T_0$ and the fluctuation of

configurational force. **(C)** Comparison of the athermal strength of NbW alloy obtained from DDD and MD simulations. **(D)** Prediction of high-temperature yield strength by the data-driven solid-solution strengthening model and experimental data. The shaded area indicates the 95% prediction interval. **(E)** Comparison between $\tau_{std}$ and size misfit parameter in various compositions. **(F)** Comparison between the CNN-predicted local fluctuational force $\tau_{std}$ and that obtained from the microelasticity model.

**Data-driven prediction of the high-temperature strength**

Current RHEAs through elemental composition tuning have shown exceptionally high-temperature strength. However, their development largely relies on trial-and-error approaches, with most studies focusing on equimolar or near-equimolar compositions(*30, 36*), leaving vast compositional spaces unexplored. Integrating multiscale theoretical models with machine learning can establish efficient composition-property relationships, accelerating the exploration of high-entropy alloy compositional spaces and enabling the design of advanced materials with superior performance. Leveraging the scaling law derived above, the critical shear stress for edge dislocation motion can be predicted. Since high-temperature strength in BCC metals is predominantly governed by edge dislocation mobility, this framework enables high-temperature strength predictions across the entire NbMoTaW alloy compositional space, beyond just equimolar systems. Fig. 6(A) illustrates the variation of $\tau_{std}$ across compositions. When ranked by ascending order of $\tau_{std}$, equimolar alloys occupy a relatively high position, consistent with their experimentally observed high-temperature strength. However, they are not the optimal compositions. Fig. 6(A) displays representative alloy compositions. For each position on the curve, the corresponding composition and its 100 nearest compositional neighbors are statistically analyzed. The shaded regions indicate the $10^{th}$-$90^{th}$ percentile ranges of elemental concentrations within these neighboring compositions. Notably, increasing Nb and W concentrations can yield properties surpassing those of equimolar alloys. Fig. 6(B)-(E) further explore compositional variations around the equimolar baseline while fixing the concentration of one element. Non-equimolar alloys exhibit enhanced high-temperature strength. This compositional exploration strategy can also be extended to other BCC alloy systems. By combining suitable interatomic potentials with the "dislocation genome" method, the microscale descriptor of $\tau_{std}$ can be derived for any BCC alloy.

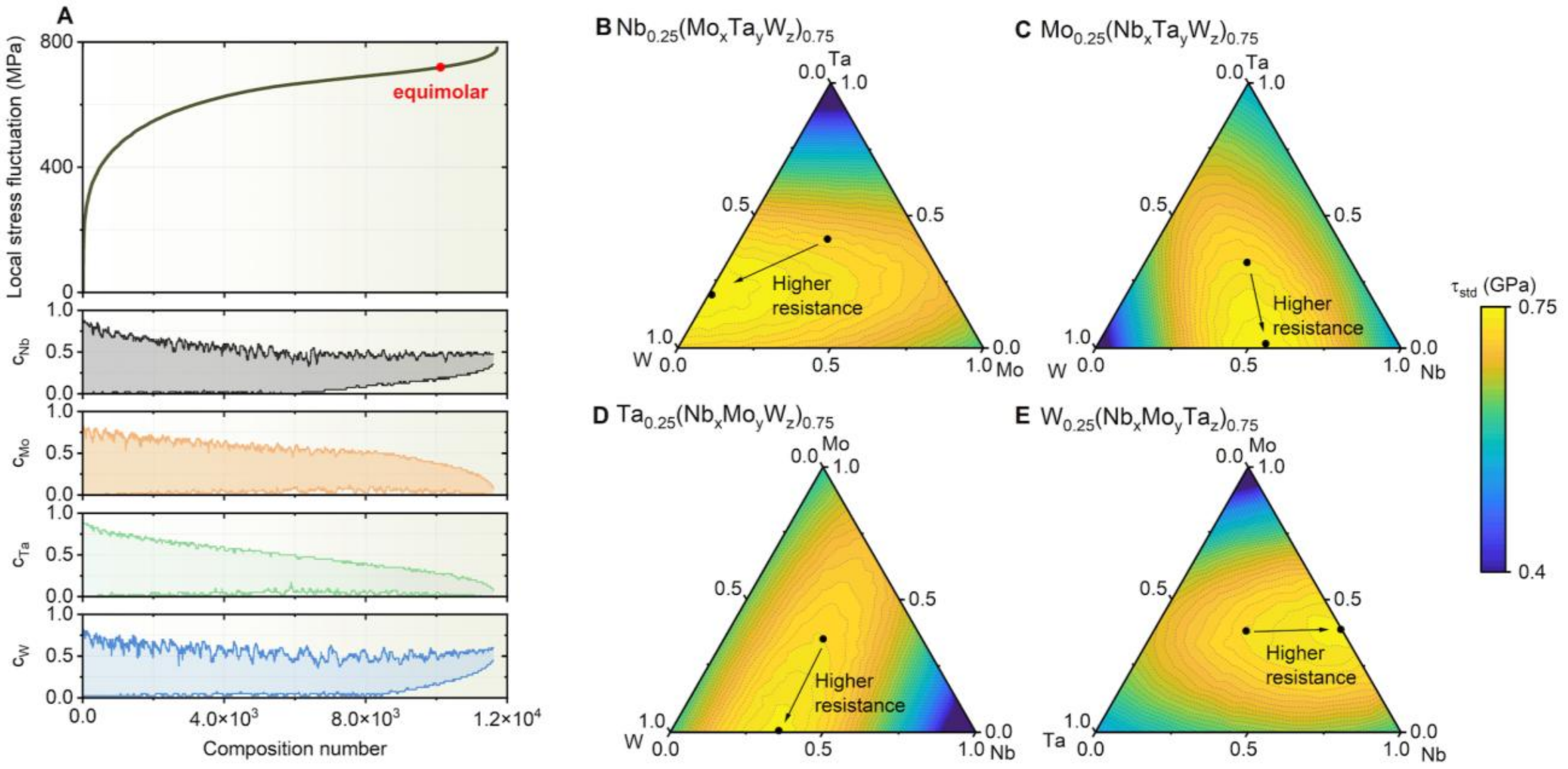

**Fig. 6 Data-driven prediction of the high-temperature strength. (A)** Theory prediction for $\tau_{std}$, which is a good indicator for the high temperature strength. Computation is performed for about 11000 alloys in the NbMoTaW compositional space. The compositions are sorted with regard to $\tau_{std}$. The shaded area of the subgraphs indicates the concentration between 10$^{th}$ and 90$^{th}$ percentiles of the neighboring 100 element contents. **(B)-(E)** Ternary graph of $\tau_{std}$ around the equimolar composition. The larger strength than that of the equimolar alloy is predicted to be exhibited in non-equimolar alloys.

## Strength and ductility prediction in NbMoTaW composition space

Based on the energy landscape for edge dislocation motion and dislocation dynamics simulations, a composition-dependent model for the resistance to edge dislocation motion is established. This model is applied to predict the high-temperature yield strength across the entire composition space.

However, high-temperature strength is not the only factor requiring attention; room-temperature ductility is also crucial for processing, manufacturing, and practical applications. Most RHEA systems suffer from a mismatch between high-temperature strength and room-temperature toughness: compositions with excellent high-temperature strength often exhibit poor toughness at room temperature, while compositional adjustments to improve room-temperature toughness frequently led to a sacrifice in high-temperature strength.

According to previous research(*37-39*), the mobility difference between screw and edge dislocations is a key indicator related to the toughness of BCC metals. Toughness depends on the competition between plastic deformation and cleavage fracture. Good plastic deformability can prevent the rapid propagation of cracks. Dislocations are crystalline defects that carry plastic deformation, and sustained plastic deformation requires the continuous generation of dislocations from dislocation sources within the metal. Experimental observations indicate that the closer the velocities (or mobilities) of screw and edge dislocations are, the higher the efficiency of dislocation sources becomes. Conversely, a large mobility gap between the two types of dislocations leads to dislocation sources becoming inactive.

The relative mobility between screw and edge dislocations is expressed as $\alpha = v_s / v_e$. It can be estimated using the difference of activation enthalpy at 0 K, approximated as $\Delta H = \Omega_s \tau_{s0} - \Omega_e \tau_{e0}$, where $\Omega_s$ and $\Omega_e$ are the activation volumes of screw and edge dislocation, respectively. $\tau_{s0}$ and $\tau_{e0}$ are the athermal resistances of the screw and edge dislocation. The specific calculation method is provided in the Supplementary Information. Figure 7 demonstrates a multi-objective optimization combining the high-temperature yield strength prediction results and the toughness prediction results. It can be observed that within the entire composition space, some non-equimolar compositions with a large portion of Mo and W with a small addition of Nb and Ta are predicted to have a combination of high-temperature strength and room-temperature ductility.

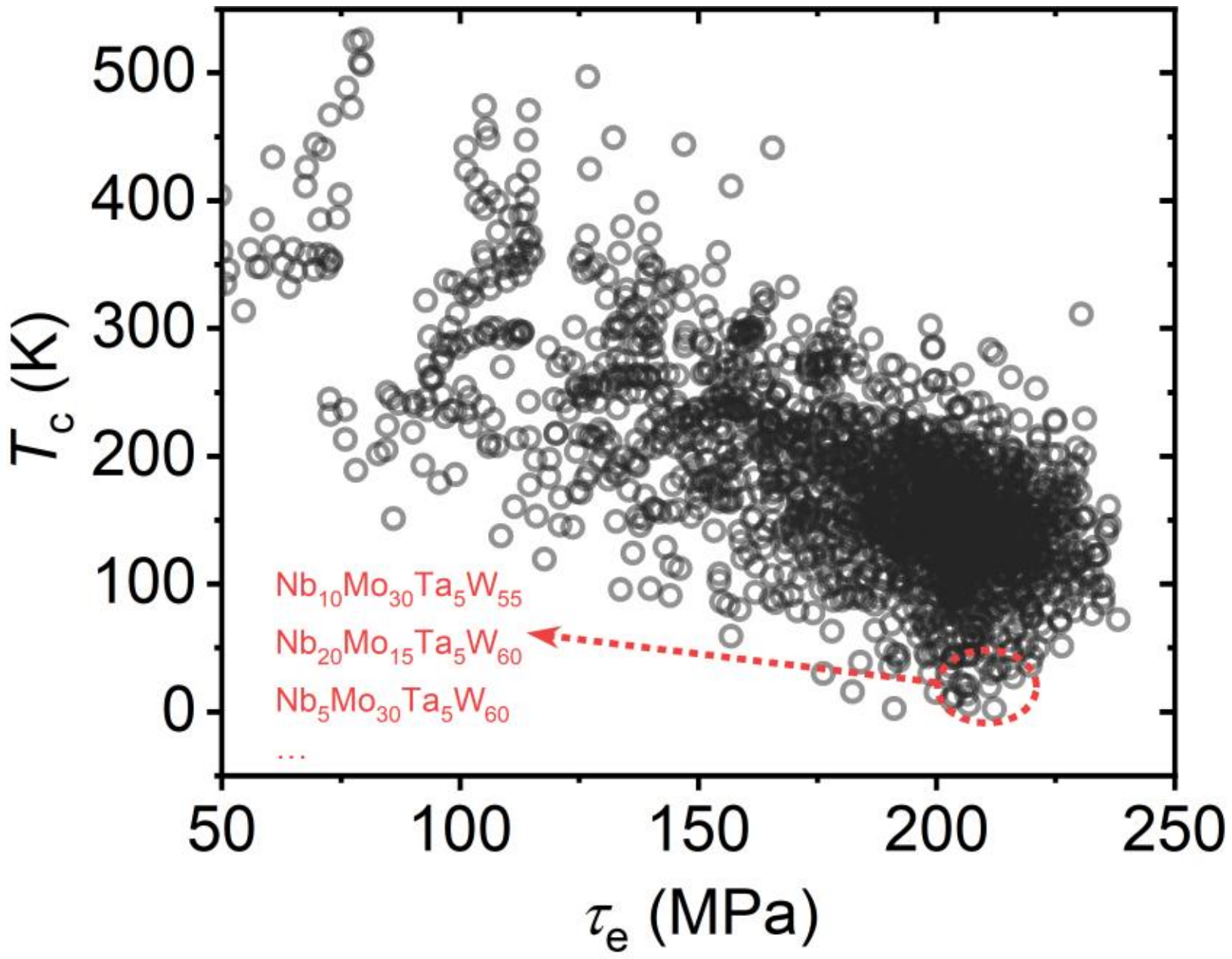

**Fig. 7 Multi-objective prediction in NbMoTaW composition space.** Characteristic temperature $T_c$ v.s. $\tau_e$ for alloys in NbMoTaW composition space. Compositions with a large portion of Mo and W with a small addition of Nb and Ta are predicted to have a combination of high-temperature strength and room-temperature ductility.

## Discussion

In this work, we establish a multiscale, data-driven framework for predicting solid-solution strengthening of edge dislocation motion in refractory BCC HEAs. Central to this framework is a "dislocation genome" model that constructs realistic energy landscapes for edge dislocation glide, from which local configurational forces and thermal fluctuations are incorporated into a discrete dislocation dynamics approach to capture mesoscale behavior in complex solid-solution environments. These simulations uncover a universal scaling relation between force fluctuations and critical shear stress, highlighting the governing role of non-equilibrium atomic configurations at dislocation cores. Building on this insight, we formulate a quantitative model that predicts the high-temperature strength of BCC alloys and integrates a ductility description based on screw-edge mobility, enabling physics-based alloy design targeting the strength–ductility balance. Coupled with high-throughput computations and experimental validation, this framework provides a generalizable paradigm to accelerate the discovery of refractory BCC HEAs and can be extended to include additional physical effects such as short-range order, magnetic interactions, and irradiation damage.

## Materials and Methods

### Embedding model for determining the energy landscape

The interaction energy $E_{\mathrm{int}}$ between an edge dislocation and solute atoms as a function of the relative distance between them is calculated by embedding the configuration into a pure metal lattice, which is termed the embedding model. Since the pure metal atoms are located far from the dislocation core, they do not contribute additional solid solution strengthening effects. Consequently, this specific atomic configuration block can be treated as isolated, effectively decoupling it from the influence of the surrounding environment. The interaction energy is then calculated using the NEB method. The details can be found in Supplementary Information.

### Neural network training in the "dislocation genome" model

A local atomic arrangement comprising 90 atoms is considered. One-hot encoding is employed to represent the atomic species within the alloy system, which consists of four elements: Nb, Mo, Ta, and W. Consequently, each atom is encoded as a four-dimensional vector. By incorporating the spatial structure of the atomic configuration, each configuration is abstracted into four matrices of size 5×16, or equivalently, viewed as a four-channel image. This representation is then processed through multiple convolutional and pooling layers. Subsequently, the output is fed into a fully connected neural network, which ultimately predicts the interaction energy between the atomic configuration and a dislocation. The neural network was implemented and trained using the PyTorch framework. The convolutional neural network has the following specific parameters: The input data dimensions are (4, 6, 15). The first convolutional layer uses a kernel size of (3, 5), produces 8 output channels, employs a stride of 1, and applies padding of (1, 2). The second convolutional layer uses a kernel size of (3, 5), produces 16 output channels, employs a stride of 1, and uses no padding. Following the convolutional layers, the data, which is flattened to 288 features, is passed to the fully connected layers. These fully connected layers have hidden layers with 128, 64, and 32 neurons, respectively, and a final output layer of size 1. The model was trained with a learning rate of $10^{-3}$, using the Mean Squared Error loss function, a batch size of 50 samples, and the Adam optimizer.

A total of 84127 data samples were generated for training, utilizing the embedding model simulated by the NEB method. This dataset comprises 27336 samples for NbW alloys, covering W concentrations ranging from 0% to 50%, and 56791 samples for equiatomic NbMoTaW alloys. During data generation for specific alloy compositions, such as the equiatomic NbMoTaW, each atomic species had an equal probability of 25% to occupy any atomic site; the exact count of each atom type within the 90-atom configuration was not strictly controlled. From the total dataset, 83000 samples were randomly selected for training, with the remainder reserved as test data. Additionally, 1000 samples for a non-equiatomic alloy composition ($Nb_{25}Mo_{25}Ta_{10}W_{40}$), not present in the training set, were generated to serve as validation data. Notably, despite the training data being predominantly centered around the equiatomic composition, the model achieved highly accurate predictions for the non-equiatomic validation set ($Nb_{25}Mo_{25}Ta_{10}W_{40}$). This demonstrates the strong scalability of the neural network developed in this work, confirming its suitability for exploring the entire compositional space of NbMoTaW alloys.

**NEB simulation of long dislocation segments**

The minimum energy path (MEP) and dislocation morphology were obtained through large-scale Nudged Elastic Band (NEB) calculations. The dimensions of the simulation supercell are 40.0 nm along the *x*-direction (parallel to the [111] crystallographic direction) and 20.7 nm along the *y*-direction (parallel to the [110] direction). Along the *z*-direction, the dislocation segment length was approximately 30 nm. Periodic boundary conditions were applied in both the *x* and *z* directions. The simulation employed 2,500 replicas to discretize the dislocation glide process over a distance of about 40 nm into segments. Each replica (image) contained approximately 1.2 million atoms. Overall, this corresponds to simulations on the order of hundreds of millions of atoms. The loading scheme for the stress-controlled NEB method is implemented using the parallel molecular dynamics simulator LAMMPS. The initial configuration of the dislocation was obtained by relaxing the supercell using the conjugate gradient method after removing atoms and embedding an edge dislocation. The final configuration was achieved by applying a shear stress to drive the dislocation motion along the *x*-direction, followed by energy minimization using the conjugate gradient method. A series of replica configurations between the initial and final states were generated via linear interpolation, with adjacent replicas coupled by spring forces. A force convergence threshold of 0.01 eV/Å was set for the calculations.

**Stochastic discrete dislocation dynamics considering configurational force**

The main differences of the DDD simulation in our work from traditional DDD simulation lie in two aspects. The first part involves calculating the forces acting on the nodes. Typical force sources include the externally applied load, line tension, and the interaction forces between dislocation segments. In this work, specifically for edge dislocations in RHEAs, an additional force component arising from the local atomic configuration is considered. This configurational force is calculated from the interaction energy constructed by the dislocation genome method, i.e.,

$$f_{\text{int}} = -\frac{\partial E_{\text{int}}}{\partial x}\ .$$

Although the configuration force represents the effect of solid-solution strengthening, it does not manifest locally solely as resistance. It overall exhibits a Gaussian distribution centered around zero. For individual dislocation segments, the probability of experiencing local resistance or propulsion is equal. However, DDD simulations demonstrate that this distribution of configuration forces does not cancel out macroscopically. Instead, it results in a net resistance effect. By integrating these configuration forces as inputs into dislocation dynamics simulations, the motion behavior of dislocations at the micrometer scale within complex solid solution atomic environments can be investigated.

The effect of thermal fluctuation, which helps the dislocation overcome energy barriers as the temperature rises, is considered by the Langevin equation. The derivation can be found in Supplementary Information. For a short dislocation segment of length $\Delta l$, the net effect of thermal fluctuation during a time interval $\Delta t$, can be expressed as

$$Q = \sqrt{\frac{2Bk_BT \cdot \Delta t}{\Delta l}} \cdot \varsigma$$

where $\varsigma \sim N(0,1)$ is a dimensionless random variable satisfying a standard normal distribution.

A dislocation length of 1000b was chosen for the simulations, with length-effect independence verified in Fig. S6. The drag coefficient *B* quantified the effect of phonon drag, which in this work is taken from results for tungsten varies with temperature. Since we focus on the thermal activation regime of dislocation motion, the value of *B* has little influence on the critical stress for dislocation motion. Therefore, variations in *B* with alloy composition are not considered (see Fig. S7). Representative dislocation dynamics show intermittent, obstacle-controlled motion and microplastic behavior preceding macroscopic yielding, with waiting times at obstacles dominating the thermally activated timescale (see Fig. S8 for details).

## References

1. J. H. Perepezko, The hotter the engine, the better. *Science* **326**, 1068-1069 (2009).

2. T. Li, T. Liu, S. Zhao, Y. Chen, J. Luan, Z. Jiao, R. O. Ritchie, L. Dai, Ultra-strong tungsten refractory high-entropy alloy via stepwise controllable coherent nanoprecipitations. *Nat. Commun.* **14**, 3006 (2023).

3. E. P. George, W. A. Curtin, C. C. Tasan, High entropy alloys: A focused review of mechanical properties and deformation mechanisms. *Acta Materialia* **188**, 435-474 (2020).

4. E. P. George, D. Raabe, R. O. Ritchie, High-entropy alloys. *Nature Reviews Materials* **4**, 515-534 (2019).

5. W. Li, D. Xie, D. Li, Y. Zhang, Y. Gao, P. K. Liaw, Mechanical behavior of high-entropy alloys. *Progress in Materials Science* **118**, 100777 (2021).

6. O. N. Senkov, D. B. Miracle, K. J. Chaput, J.-P. Couzinie, Development and exploration of refractory high entropy alloys—A review. *Journal of Materials Research* **33**, 3092-3128 (2018).

7. T. Tsuru, S. Han, S. Matsuura, Z. Chen, K. Kishida, I. Iobzenko, S. I. Rao, C. Woodward, E. P. George, H. Inui, Intrinsic factors responsible for brittle versus ductile nature of refractory high-entropy alloys. *Nat. Commun.* **15**, 1706 (2024).

8. W. Xiong, A. X. Y. Guo, S. Zhan, C.-T. Liu, S. C. Cao, Refractory high-entropy alloys: A focused review of preparation methods and properties. *Journal of Materials Science & Technology* **142**, 196-215 (2023).

9. Z. Wang, S. Chen, S. Yang, Q. Luo, Y. Jin, W. Xie, L. Zhang, Q. Li, Light-weight refractory high-entropy alloys: A comprehensive review. *Journal of Materials Science & Technology* **151**, 41-65 (2023).

10. X. Wang, L. Valdevit, P. Cao, Neural network for predicting Peierls barrier spectrum and its influence on dislocation motion. *Acta Materialia* **267**, 119696 (2024).

11. F. Wang, G. H. Balbus, S. Xu, Y. Su, J. Shin, P. F. Rottmann, K. E. Knipling, J.-C. Stinville, L. H. Mills, O. N. Senkov, I. J. Beyerlein, T. M. Pollock, D. S. Gianola, Multiplicity of dislocation pathways in a refractory multiprincipal element alloy. *Science* **370**, 95-101 (2020).

12. D. Utt, S. Lee, Y. Xing, H. Jeong, A. Stukowski, S. H. Oh, G. Dehm, K. Albe, The origin of jerky dislocation motion in high-entropy alloys. *Nature Communications* **13**, 4777 (2022).

13. L. Zhang, Y. Xiang, J. Han, D. J. Srolovitz, The effect of randomness on the strength of high-entropy alloys. *Acta Materialia* **166**, 424-434 (2019).

14. T. Jiang, Y. Xiang, L. Zhang, Stochastic Peierls--Nabarro model for dislocations in high entropy alloys. *SIAM Journal on Applied Mathematics* **80**, 2496-2517 (2020).

15. F. Shuang, L. Laurenti, P. Dey, Standard deviation in maximum restoring force controls the intrinsic strength of face-centered cubic multi-principal element alloys. *Acta Materialia* **282**, 120508 (2025).

16. W. Li, S. Lyu, Y. Chen, A. H. W. Ngan, Fluctuations in local shear-fault energy produce unique and dominating strengthening in metastable complex concentrated alloys. *Proceedings of the National Academy of Sciences U.S.A* **120**, e2209188120 (2023).

17. Q.-J. Li, H. Sheng, E. Ma, Strengthening in multi-principal element alloys with local-chemical-order roughened dislocation pathways. *Nature Communications* **10**, 3563 (2019).

18. S. Yin, Y. Zuo, A. Abu-Odeh, H. Zheng, X.-G. Li, J. Ding, S. P. Ong, M. Asta, R. O. Ritchie, Atomistic simulations of dislocation mobility in refractory high-entropy alloys and the effect of chemical short-range order. *Nature Communications* **12**, 4873 (2021).

19. S. Lyu, W. Li, Y. Xia, Y. Chen, A. H. W. Ngan, Effects of chemical randomness on strength contributors and dislocation behaviors in a bcc multiprincipal element alloy. *Physical Review Materials* **7**, 073602 (2023).

20. A. A. Catal, E. Bedir, R. Yilmaz, M. A. Swider, C. Lee, O. El-Atwani, H. J. Maier, H. C. Ozdemir, D. Canadinc, Machine learning assisted design of novel refractory high entropy alloys with enhanced mechanical properties. *Computational Materials Science* **231**, 112612 (2024).

21. S. Zhao, J. Li, W. Liao, R. Yuan, Machine learning-guided plasticity model in refractory high-entropy alloys. *MGE Advances* **3**, e70022 (2025).

22. B. Xing, T. J. Rupert, X. Pan, P. Cao, Neural network kinetics for exploring diffusion multiplicity and chemical ordering in compositionally complex materials. *Nature Communications* **15**, 3879 (2024).

23. S. Kumar, V. Jindal, Integrating machine learning and DFT for hardness prediction in high-entropy alloys. *MRS Communications*, (2025).

24. C. Tandoc, Y.-J. Hu, L. Qi, P. K. Liaw, Mining of lattice distortion, strength, and intrinsic ductility of refractory high entropy alloys. *npj Computational Materials* **9**, 53 (2023).

25. F. Zhao, W. Liu, Y. Zhang, H. Duan, The hierarchical energy landscape of edge dislocation glide in refractory high-entropy alloys. *Journal of the Mechanics and Physics of Solids* **193**, 105887 (2024).

26. G. Henkelman, B. P. Uberuaga, H. Jónsson, A climbing image nudged elastic band method for finding saddle points and minimum energy paths. *The Journal of chemical physics* **113**, 9901-9904 (2000).

27. K. J. Wiese, Theory and experiments for disordered elastic manifolds, depinning, avalanches, and sandpiles. *Reports on Progress in Physics* **85**, 086502 (2022).

28. D. A. Koss, Deformation of Nb-W single crystals: Athermal solid solution strengthening. *Metallurgical Transactions* **2**, 245-248 (1971).

29. J. R. Slining, D. A. Koss, Solid Solution strengthening of high purity niobium alloys. *Metallurgical Transactions* **4**, 1261-1264 (1973).

30. O. N. Senkov, G. B. Wilks, J. M. Scott, D. B. Miracle, Mechanical properties of $Nb_{25}Mo_{25}Ta_{25}W_{25}$ and $V_{20}Nb_{20}Mo_{20}Ta_{20}W_{20}$ refractory high entropy alloys. *Intermetallics* **19**, 698-706 (2011).

31. R. W. Buckman, New applications for tantalum and tantalum alloys. *The Journal of The Minerals, Metals & Materials Society* **52**, 40-41 (2000).

32. C. D. Statham, D. A. Koss, J. W. Christian, The thermally activated deformation of niobium-molybdenum and niobium-rhenium alloy single crystals. *Philosophical Magazine A* **26**, 1089-1103 (1972).

33. K. Joshi, P. Kumar, Strength Behavior of Niobium-Based Refractory Systems. *The Journal of The Minerals, Metals & Materials Society* **76**, 6277-6301 (2024).

34. G. Ma, Z. Wei, G. Wu, X. Mao, The microstructure and strength of a tantalum alloy: Influence of temperature. *Materials Science and Engineering: A* **880**, 145312 (2023).

35. P.-A. Geslin, D. Rodney, Microelasticity model of random alloys. Part I: mean square displacements and stresses. *Journal of the Mechanics and Physics of Solids* **153**, 104479 (2021).

36. F. Maresca, W. A. Curtin, Mechanistic origin of high strength in refractory BCC high entropy alloys up to 1900K. *Acta Materialia* **182**, 235-249 (2020).

37. Y. Lu, Y.-H. Zhang, E. Ma, W.-Z. Han, Relative mobility of screw versus edge dislocations controls the ductile-to-brittle transition in metals. *Proceedings of the National Academy of Sciences* **118**, e2110596118 (2021).

38. Y.-H. Zhang, E. Ma, J. Sun, W.-Z. Han, A unified model for ductile-to-brittle transition in body-centered cubic metals. *Journal of Materials Science & Technology* **141**, 193-198 (2023).

39. Y.-H. Zhang, W.-Z. Han, Dislocation source efficiency as a governing factor in the ductile-to-brittle transition of metals. *Journal of Materials Science & Technology* **229**, 173-176 (2025).

## Acknowledgements

The authors gratefully acknowledge the computational resources provided by the Computing Centers in Xi'an.

**Funding:**

This work was supported by the National Natural Science Foundation of China Project No. 12588301, 12472103, and Science Challenge Project No. TZ2025016.

**Author contributions:**

Conceptualization: FZ, YZ

Methodology: FZ,YZ

Investigation: JZ, YZ, HD

Software: FZ, YZ

Validation: FZ

Formal analysis: FZ, YZ

Writing—original draft: FZ

Writing—review & editing: JZ, YZ, HD

Supervision: HD

Project administration: HD

Funding acquisition: HD

**Competing interests:**

Authors declare that they have no competing interests.

**Data and code availability:**

All data needed to evaluate the conclusions in the paper are present in the paper and/or the Supplementary Materials.